# Sub-angstrom Non-invasive Imaging of Atomic Arrangement in 2D Hybrid Perovskites

Mykola Telychko[1,†], Shayan Edalatmanesh[2,3,†], Kai Leng[1,†], Ibrahim Abdelwahab[1,4], Na Guo[5], Chun Zhang[5], Jesús I. Mendieta-Moreno[2], Matyas Nachtigall[2], Jing Li[4], Kian Ping Loh[1]*, Pavel Jelínek[2,3]*, Jiong Lu[1,4]*

[1]Department of Chemistry, National University of Singapore, 3 Science Drive 3, Singapore 117543, Singapore

[2]Institute of Physics, The Czech Academy of Sciences, 162 00 Prague, Czech Republic

[3]Regional Centre of Advanced Technologies and Materials, Palacký University, 78371 Olomouc, Czech Republic

[4]Centre for Advanced 2D Materials (CA2DM), National University of Singapore, 6 Science Drive 2, Singapore 117546, Singapore

[5]Department of Physics, National University of Singapore, Singapore, Blk S12, Science Drive 3, Singapore 117551

†These authors contributed equally to this work: M. Telychko, S. Edalatmanesh, K. Leng

★Corresponding authors. Email: chmluj@nus.edu.sg (J. Lu); jelinekp@fzu.cz (P. J.); chmlohkp@nus.edu.sg (K. P. L.)

# Abstract

Non-invasive imaging of the atomic arrangement in two-dimensional (2D) Ruddlesden–Popper hybrid Perovskites (RPPs), as well as understanding the related effects is challenging, due to the insulating nature and softness of the organic layers which also obscure the underlying inorganic lattice. Here, we demonstrate a sub-angstrom resolution imaging of both soft organic layers and inorganic framework in a prototypical 2D lead-halide RPP crystal *via* combined tip-functionalized Scanning Tunneling Microscopy (STM) and non-contact Atomic Force Microscope (ncAFM) corroborated by theoretical simulations. STM measurements unveil the atomic reconstruction of the inorganic lead-halide lattice and overall twin-domain composition of the RPP crystal, while ncAFM measurements with a CO-tip enable non-perturbative visualization of the cooperative reordering of surface organic cations driven by their hydrogen bonding interactions with the inorganic lattice. Moreover, such a joint technique also allows for the atomic-scale imaging of the electrostatic potential variation across the twin-domain walls, revealing alternating quasi-one-dimensional (1D) electron and hole-channels at neighboring twin-boundaries, which may influence in-plane exciton transport and dissociation.

**Introduction**

The 2D RPPs offer a remarkably rich material platform for optoelectronic device applications, for which the excitonic properties are closely linked to their quantum well structures consisting of soft insulating organic layers sandwiched between conducting inorganic lead-halide frameworks. The presence of additional soft insulating organic layers in 2D RPPs versus three-dimensional (3D) counterparts introduces the two-dimensionality, the emergence of many quantum phenomena[1-4], and also leads to significantly enhanced photo- and chemical stability[5,6], and tunable optoelectronic properties[7-11]. Such a unique dielectric and quantum confinement effects establish RPPs as a promising class of materials for next-generation optoelectronic applications[12-16]. Furthermore, recent photoluminescence (PL) studies demonstrate that organic cations in RPPs and polarizable inorganic lattice are prone to cooperative structural relaxations under external perturbations including strain and light exposure, and thus dramatically alter the optoelectronic response of RPPs[17]. Analogously to the 3D counterparts, the structural relaxations of the inorganic lattice of 2D RPPs may lead to the emergence of the various ferroelastic domains and their associated twin-boundaries, which have not been hitherto studied at atomic-scale neither in 3D nor in 2D hybrid perovskites.

Superior advantages of RPPs for a wide range of photovoltaic and optoelectronic applications, are further exemplified by a number of recently uncovered phenomena related to exciton transport and recombination in these materials. For instance, the most recent experimental breakthroughs show that RPP crystals render a remarkably long-range exciton diffusion (hundreds of nanometers) and reduced rates of exciton recombination[18,19]. Furthermore, it has been demonstrated that charge transport is largely determined by intrinsic crystalline structure of the RPP's, pronounced exciton-polaronic effects[20,21] and peculiarities of the energy landscape[22]. Nevertheless, fundamental understanding of these charge transport phenomena in RPP is not complete, as it requires the microscopic knowledge of the structural properties of RPP, which remains elusive to date.

Future progress in this field hinges on the atomic scale understanding of the dynamic structural re-ordering and cooperative lattice relaxation in the crystal that impact their electronic, optoelectronic and excitonic properties. Therefore, a real-space non-invasive atomic imaging of both, top organic layers and underlying inorganic lattice in a 3D fashion is highly desired. Unfortunately, the insulating nature and softness of organic layers as well as "buried" inorganic framework render atomically-resolved imaging of the RPPs a grand challenge, beyond the capabilities of the current state-of-the-art imaging techniques including both, Scanning Tunneling Microscopy (STM) and Scanning Transmission Electron Microscopy (STEM). Although tunneling through insulating organic layers can be exploited to resolve the atomic lattice of inorganic framework *via* STM, the soft organic chains can be easily excited *via* inelastic tunneling process, leading to their relocation and disordering of BA$^+$ layer. Alternatively, beam-efficient STEM imaging techniques have been developed to image inorganic lattices of the 3D organic-inorganic perovskites[23-25]. However, the application of STEM to image 2D organic-inorganic RPPs causes the structural damage to such beam-sensitive RPPs due to the collisions of the soft organic layers with the energetic electron beam[26].

Recent advances in a tuning-fork (Qplus) based ncAFM imaging technique with a carbon monoxide (CO)-functionalized tip, have established this technique as a powerful tool for atomically-resolved studies[27]. It has been demonstrated that ncAFM with a judiciously-decorated tip offers an extraordinary sub-angstrom (sub-Å) resolution imaging (< 1 Å) of solid surface[28], organic materials[29-31] and even non-perturbative imaging of weakly-bonded systems such as water clusters[32,33]. Therefore, Qplus-ncAFM technique potentially acts as an ideal tool for non-invasive sub-Å scale imaging of insulating organic space layers in RPPs. To this end, we employed a low-temperature STM and ncAFM imaging to resolve atomic structures of both inorganic framework and organic layers in the prototypical hybrid lead-halide RPP crystal. The STM imaging resolves the reconstruction of the inorganic octahedral framework and unveils the twin-domain composition of the RPPs, whilst ncAFM imaging with a CO-functionalized tip enables a non-invasive visualization of the cooperative reconstruction of the on-surface $BA^+$ cations (Fig. 1a), presented by a well-ordered array of the cation pairs. The structural relaxation of the organic layers is interlocked with the deformation of the inorganic lattice through hydrogen bonding, corroborated by density functional theory (DFT) calculations. The lattice deformation leads to the formation of ferroelastic domains stitched *via* quasi-one-dimensional twin-boundaries with spatial extension over hundreds of nanometers. We demonstrate the atomic-scale imaging of the electrostatic potential variation across the twin-domain walls, revealing alternating quasi-one-dimensional (1D) electron and hole channels at neighboring twin-boundaries, which sheds new light on the mechanisms of the efficient separation of photoexcited electron–hole pairs and exciton transport in 2D RPPs.

**Results and Discussion**

**Exfoliation of 2D RPPs atomic layers for the STM measurements.** We have selected the *n* = 4 homologue of the lead-iodine (Pb-I) RPP family for the STM imaging. The RPP family is described by a general chemical formula $(CH_3(CH_2)_3NH_3)_2(CH_3NH_3)_{n-1}Pb_nI_{3n+1}$., where *n* reflects the number of inorganic octahedral sheets (typically $n$ = 1 to 4) sandwiched by butylammonium ($CH_3(CH_2)_3NH_3$) cations (denoted as $BA^+$). In analogy to the 3D counterpart, inorganic RPP's framework consists of the corner-sharing $[PbI_6]$ octahedral cages. Furthermore, short polar methylammonium ($CH_3NH_3$) (denoted as $MA^+$) molecules reside in the intra-octahedral space to balance negative electrostatic charge of the $[PbI]^-$ lattice. The presence of insulating $BA^+$ cations in the RPPs significantly reduces its out-of-plane conductivity and precludes STM study of bulk samples.

To overcome this issue, we mechanically exfoliated a bulk $n$=4 RPP crystals onto Au(100) substrate to produce monolayer and few-layer flakes for a combined STM/ncAFM measurements at 4.5 K. The layer-dependent optical contrast (Fig. 1a) facilitates the tip positioning over RPP flakes with any desired thickness. The mechanical exfoliation generally produces $BA^+$– terminated single crystal flakes. Representative STM images of a few-layered RPP flake acquired at positive (Fig. 1b) and negative (Supplementary Fig. S1b) sample bias voltages ($V_S$) reveal a periodic dimer-like pattern and disordered structure respectively. The dimer-like pattern resembles the one acquired on the surfaces of 3D hybrid lead-halide perovskites, including $MA^+PbBr_3$[34-36], $MA^+PbI_3$[37-39] and $CsPbBr_3$[40], which has been ascribed to a pair of halogen atoms located at apices of neighboring $[PbI_6]$ octahedra, arising from the ferroelectric-like organizations of the inner $MA^+$ cations. Each individual dimer observed here also contains a pair of protrusions, corresponding to a pair of the topmost iodine atoms (I-dimers) located at apices of the proximate $[PbI_6]$ octahedra [39]. The exclusive presence of the dimer-like pattern in RPP is in stark contrast to case of 3D hybrid perovskites, which also reveal the zigzag-like STM patterns, associated with antiferroelectric-like organizations of the inner $MA^+$ cations [39]. Such a different behavior is presumably attributed to peculiar 2D nature of the RPP perovskites that favors the sole ferroelectric-like arrangement of the inner MA$^+$ chains, leading to the emergence of extended ferroelastic domains.

The perceived bias-dependent STM contrast is closely related to the peculiar electronic structure of the RPP. The calculated partial density of electronic states (PDOS) reveals that the conduction band (CB) of the RPP crystal is composed presumably of 6*s* Pb and 5*p* I electronic states (Supplementary Fig. S5 and Supplementary Note S2). Therefore, the tunneling current is dominated by the topmost I atoms leading to the emergence of dimer-like STM pattern at positive bias voltage (see Supplementary Note 1). In contrast, valence band (VB) is predominantly formed by 5*p* I states, but also contains a non-negligible contribution of the BA$^+$ induced states, which allows for the visualization of the on-surface BA$^+$ cations at negative sample bias voltages. Notably, the flexible BA$^+$ cations can be easily excited by tunneling electrons, appearing as fuzzy features in the STM images (Supplementary Fig. S1).

**ncAFM measurements of the RPPs.** To circumvent the challenge in the imaging of soft BA$^+$ layers, we employed the constant-height ncAFM imaging with a CO-terminated tip, allowing for a non-invasive imaging of BA$^+$ molecules on the RPP surface. The frequency shift (*Δf*) contrast in the constant-height ncAFM images is associated with the spatial variation of an electrostatic, dispersive and repulsive forces acting between CO-terminated tip and BA$^+$ cations in the Pauli repulsion regime[31]. To eliminate the possible perturbations of BA$^+$ chains by tunneling process, tip was repositioned (typically > 50 nm) over the intact RPP sample area for the subsequent ncAFM imaging at zero bias voltage.

A representative high-resolution ncAFM image of the RPP surface (Fig. 2a) reveals "arrow-like" features arranged in a square-shaped array (see the additional large-scale ncAFM image in

Supplementary Fig. S2). Here, each "arrow-like" feature represents a pair of apical methyl (-$CH_3$) groups of the two adjacent $BA^+$ molecules (denoted as $BA^+$– pair). Combined with subsequent STM imaging acquired over the same sample area, the location of the adsorbed (on-surface) $BA^+$ cations with respect to the underlying octahedral $[PbI]^-$ lattice can be determined unambiguously. Superimposing the underlying inorganic lattice determined from STM image (Fig. 2C), over the corresponding ncAFM image (Fig. 2b) reveals that $BA^+$– pairs are located in-between apical I atoms and are aligned along [100] and [010] directions of the $[PbI]^-$ lattice. These findings further underpin the extraordinary capability of the combined STM/ncAFM measurements to unveil the geometry of both, organic $BA^+$ arrays and the underlying inorganic octahedral lattice through non-invasive imaging of an RPP structure in a quasi-3D fashion.

In order to understand the ncAFM contrast of the $BA^+$ – pairs associated with their peculiar conformations, we acquired frequency shift versus tip-sample distance ($(\Delta f(\Delta z))$) curves over several characteristic sites (Fig. 2e), including over and between protruding methyl groups of the $BA^+$ chains, indicated by color-coded arrows in the inset of Fig. 2e. The $\Delta f(z)$ curves exhibit a minimum in the vicinity of $\Delta z \approx 1$Å, at which the attractive and repulsive tip-sample forces are balanced[41]. The $\Delta f(\Delta z)$ curves measured over two paired $BA^+$ cations reveal a height variation of ~ 44 pm, estimated as the difference between the minima of the respective $\Delta f(z)$ curves. As shown in Fig. 2d, we also acquired a set of constant-height $\Delta f$ images over an individual $BA^+$– pair at different tip-sample distances (also Supplementary Fig. S2b-f). The constant-height $\Delta f$ image taken at relatively large $\Delta z$ shows a depression over the right $BA^+$ molecule due to an attractive tip-sample interaction. The left $BA^+$ chain has a larger surface protrusion of ~ 44 pm compared to the right $BA^+$ chain, leading to its brighter appearance arising from a stronger repulsive tip-sample interaction. A further decrease of $\Delta z$ results in a gradual emergence of the sharp feature associated with H atoms at -CH$_3$ groups of BA$^+$ molecule.

**Mechanism of the BA$^+$ pairing and BA$^+$ interaction with I-dimers.** The aforementioned findings suggest that BA$^+$ cations pairing and their cooperative periodic arrangement with respect to the octahedral [PbI]$^-$ lattice is likely driven by the short-range electrostatic interactions between the inorganic lattice and BA$^+$ cations. To gain a deeper insight into the origin of the unique arrangement of the BA$^+$ cations, we performed large-scale DFT+vdW calculations of RPP in the orthorhombic configuration[42,43]. The DFT-relaxed atomic structure (Fig. 3d,e) reveals that the organic cations (MA$^+$ and BA$^+$) form electrostatic hydrogen-like bonds (denoted as H-bonds)[44-45] through their positively charged amine terminal groups (-NH$_3$) with the negatively charged I atoms from the inorganic cage. This interaction is not only responsible for the asymmetric arrangement of BA$^+$ molecules, but also causes the structural deformation of the inorganic lattice, manifested by distortions of [PbI$_6$] cages not only in the *xy*-plane, but also along the *z*-axis (Fig. 3f). This also induces a significant surface relaxation of the organic cations, manifested by formation of the BA$^+$-pairs. In addition, the simulated CO-tip ncAFM images of DFT-relaxed RPP model (Fig. 3b,c) using the Probe Particle (denoted as PP) SPM Model[46-48] reproduce the contrast observed in the experimental ncAFM images reasonably well (Fig. 3A). Furthermore, the PP-ncAFM images, simulated at various $\Delta z$ (Supplementary Fig. S3), fully corroborate the experimental $\Delta z$-dependent evolution of the ncAFM contrast of BA$^+$ pair (Fig. 2e), which further attests the validity of the uncovered atomic RPP structure.

**The manipulation of BA$^+$ pairs.** Interestingly, along with aforementioned most abundant "arrow-like" conformation of the BA$^+$ pairs (denoted as type-I), ncAFM images also occasionally reveal the presence of another conformation type of the $BA^+$ pairs, which assume a "T-like" form (denoted as type-II). These conformations are interconvertible *via* STM imaging at elevated bias voltages. As shown in Fig. 3a, a sequence of ncAFM images of the same surface area capture the tip-assisted transformation of the $BA^+$– pairs, from type-I to type-II (additional data in Supplementary Fig. S4).

To investigate the mechanism behind the evolution of a $BA^+$ pair from type-I to type-II configuration, we performed a computational search for the other possible configurations of the $BA^+$ molecules. Indeed, we found another stable configuration (Fig. 3e), which differs in the total energy only by tens of meV with respect to the original type-I configuration. One $BA^+$ cation in this new configuration (i.e. type-II) preserves its position, while the upper part of another $BA^+$ cation tilts along the [001] axis. The corresponding simulated PP-ncAFM image of this new configuration (Fig. 3c) also matches well to the experimental one. Throughout this transformation, inorganic cage shows a negligible distortion since the amine groups of the $BA^+$ chains still remain at their original locations *via* H-bonding with I atoms. Thus, applying bias voltage accompanied with the tunneling process locally modifies the orientation of $BA^+$ chains, without disruption of the ionic lattice.

**Origin of the twin–domain composition of the RPP crystal.** Our large-scale STM imaging unambiguously reveals that the RPP surface exclusively consists of the alternating ferroelastic[49] domains (color-coded as green and red in Fig. 4a) and twin–boundaries, that are nearly parallel–aligned along [100] direction. The I–dimers in adjacent domains are found to be orientated along the [101] and [10-1] lattice directions. Such a twin–domain composition of the RPP crystal is attributed to the cooperative ferroelectric-like alignment of the inner $MA^+$ dipoles within each octahedral sheet. Specifically, on the basis of the fact that $n$=4 RPP hosts an odd number of $MA^+$ layers (three), the individual RPP domain is expected to attain a net non-zero polarization associated with the cooperative alignment of the $MA^+$ chains. Therefore, the twin–domains form as a result of the energetically favorable antiferroelectric-like net polarization of the $MA^+$ layers residing in adjacent domains. The twin-domain composition of the RPP crystalline flakes, with quasi-one-dimensional twin-boundaries extended over hundreds of nanometers has not been reported to date. Despite the observation of twin-domain boundaries in 3D $MA^+PbI_3$ films using X-ray diffraction, TEM and piezoresponse force microscopy imaging techniques, [49-53] as well as the STM studies of 3D hybrid perovskites, [34-38] neither one reveals the formation of ferroelastic domains, analogous to one we observe in 2D RPP.

Such a twin-domain composition of the RPP consists of two types of domain boundaries, namely "head-to-head" (Fig. 4b) and "tail-to-tail" (Fig. 4d), which were predicted to host positive and negative electrostatic potential respectively[54-55]. To verify this hypothesis, we conducted Kelvin probe force microscopy measurements (KPFM) (see Methods and Supplementary Note 3) to probe the local contact potential difference (LCPD), which can be associated with the spatial variation of the local work function[56] across the domain walls. LCPD maps unambiguously show a lower and higher LCPD value over "head-to-head" (Fig. 4c) and "tail-to-tail" (Fig. 4e) twin–boundaries, respectively.

Such a characteristic LCPD variation across the twin-boundaries has been consistently observed by multiple sessions of KPFM measurements, which suggest the presence of a positive (negative) electrostatic potential in the vicinity of "head-to-head" ("tail-to-tail") twin-boundaries. It needs to be noted that absolute magnitude of the LCPD contrast does not reflect the magnitude of the electrostatic potential, since CPD value strongly depends on the structure of tip apex and KPFM acquisition setpoint.[56] To illustrate this, we present KPFM measurements acquired using the same tip-apex at varying tip-sample distances in Fig. S6, which reveals a remarkable change of the CPD value upon a decrease of the tip-sample distance.

The contrast LCPD maps is in good agreement with the landscape of the electrostatic potential over "head-to-head" ("tail-to-tail") twin-boundaries calculated for a twin–boundaries slab model depicted in Fig. 4f. Here, to decipher the origin of the LCPD signal across the domain boundaries, we designed the slab model consisting of 3600 atoms including both types of twin–boundaries separated by a buffer level ensuring sufficient separation of the twin-boundaries. We carried out large scale total energy DFT-vdW calculations to obtain the fully optimized atomic structure of $n=4$ RPP. We then calculated individual dipole moments of individual $BA^+$ and $MA^+$ molecules for constructing an effective lattice dipole model (see Supplementary Note 4 for more details). This model allows us to decipher the impact of the individual molecular $BA^+$ and $MA^+$ layers on the resulting electrostatic potential. We found that the outermost $MA^+$ molecular layer plays the decisive role on the character of the electrostatic potential across the domain walls.

We note that intrinsic ferroelastic-like composition of the 2D RPP crystals, uncovered by our STM/ncAFM studies, can be a key towards understanding their outstanding photovoltaic characteristics and occurrence of a number of intriguing optical properties unveiled recently[17-22]. Specifically, recent temperature-dependent PL studies reveal an emergence of shallow electronic states in the $n=4$ $(BA^+)_2(MA^+)_{n-1}Pb_nI_{3n+1}$ RPP with an increased emission lifetime. These electronic states were previously ascribed to the presence of local domains and static disorder in the crystal that emerged at low-temperature[22], although their structure has not been verified. Our findings not only provide a new quantitative insight into nanoscale domain composition of the RPPs, but also suggest that low-energy electronic states revealed by PL spectra are likely associated with the domain structures observed here.

Furthermore, the polarized electrostatic potential across twin-boundaries, revealed by our KPFM studies and corroborated by DFT calculations, is expected to facilitate the separation of electron-hole pairs at "head-to-head" and "tail-to-tail" twin-boundaries hosting opposite electrostatic potential, thus leading to reduced rates of electron-hole recombination in RPPs. Analogous to this, the separation of electron and hole pairs at ferroelastic domains boundaries have been proposed to result in a reduced rate of exciton recombination in 3D hybrid perovskites[57-59]. However, the electrostatic potential of these domains boundaries has not been quantitatively probed to date.

Lastly, owing to their one-dimensional nature (hundreds of nanometers), twin-boundaries can be viewed as channels for a long-distance exciton propagation, which is directly corroborated by a recently reported phenomenon of the long-range (hundreds of nm) exciton funneling in RPPs[18,19,22]. It is therefore envisaged that controlling the orientation of the twin-boundaries *via* external stimuli[58,59,17] can be a plausible strategy to enhance the performance of the RPP-based photovoltaic and optoelectronic devices.

**Conclusion**

We have achieved sub-Å resolution non-invasive imaging of the atomic structure of a prototypical RPP hybrid perovskite using STM and tip-functionalized ncAFM. The STM imaging resolves the atomic reconstruction of the underlying inorganic lead-halide lattice, whereas ncAFM measurement with a CO-tip allows for a non-perturbative visualization of the cooperative reordering of surface organic cations. In addition, combined STM/ncAFM measurements also reveal the existence of domains in the RPP crystal, which arise from the cooperative alignment of interior polar $MA^+$ chains and a synergetic distortion of the $[PbI]^-$ lattice. Our results suggest that the reorganization of surface lattices and existence of charged domain boundaries may have important implications for the optoelectronic properties of thin 2D perovskites films. The non-destructive imaging of the microstructures and ferroelastic domains in the hybrid perovskites with sub-Å resolution provides unprecedented insights into their nanoscale properties and outstanding optoelectronic device performance.

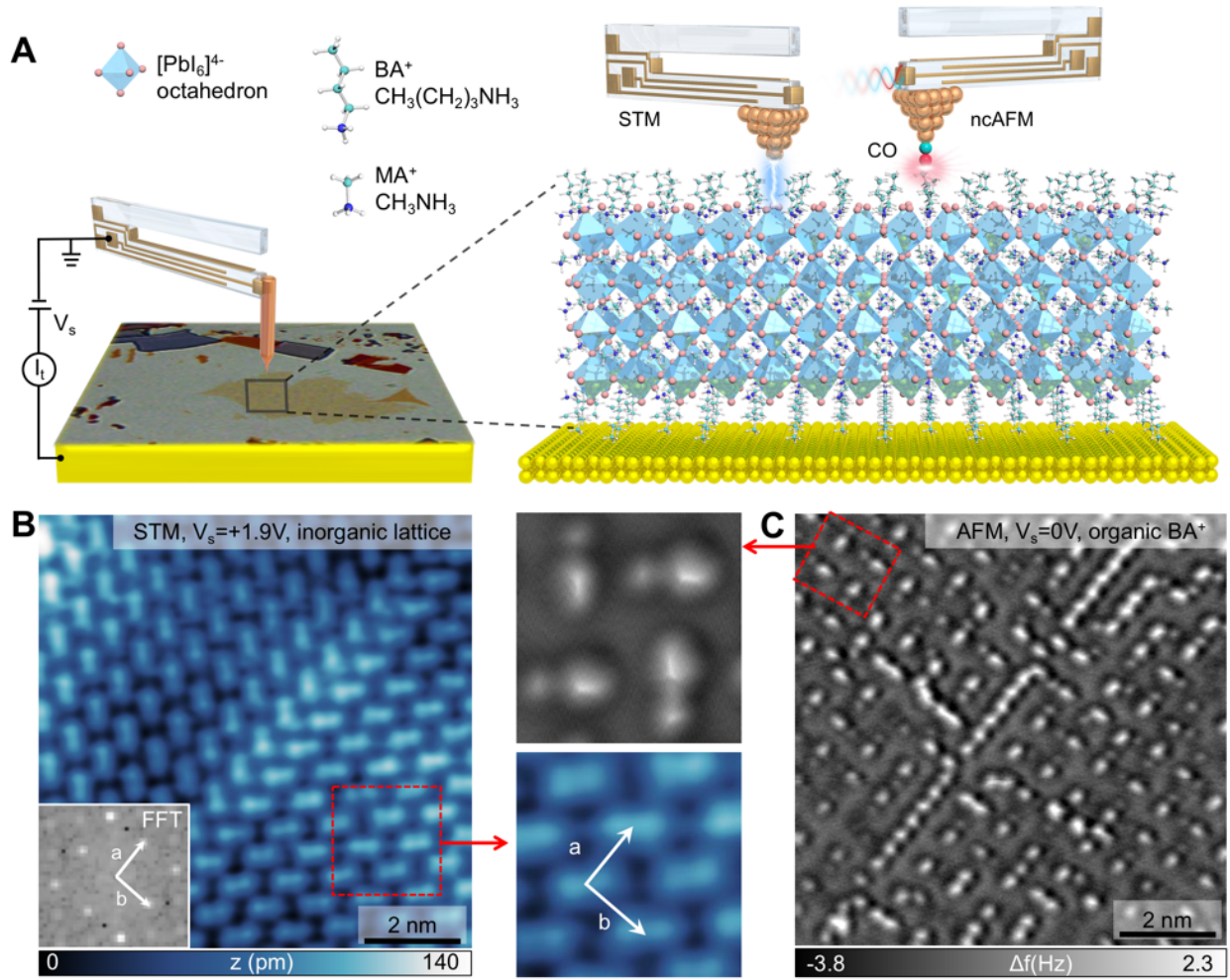

**Fig. 1. qPlus based STM and ncAFM imaging of the RPP surface**. (**A**) Schematics showing a combined STM and ncAFM imaging of the RPP surface using a tuning fork - based Qplus sensor. Atomic layers of the RPP crystals are obtained by a mechanical exfoliation and then transferred onto conducting Au substrate (optical image on left). (**B**) STM image of RPP acquired at positive sample bias voltage ($V_s$= +1.9 V). (**C**) ncAFM image collected over the same surface area. ncAFM image was acquired in constant-height mode, at a tip-sample distance of $\Delta z$= +100 pm with respect to an original set-point of $V_s$ = 2 V and $I$ = 15 pA.

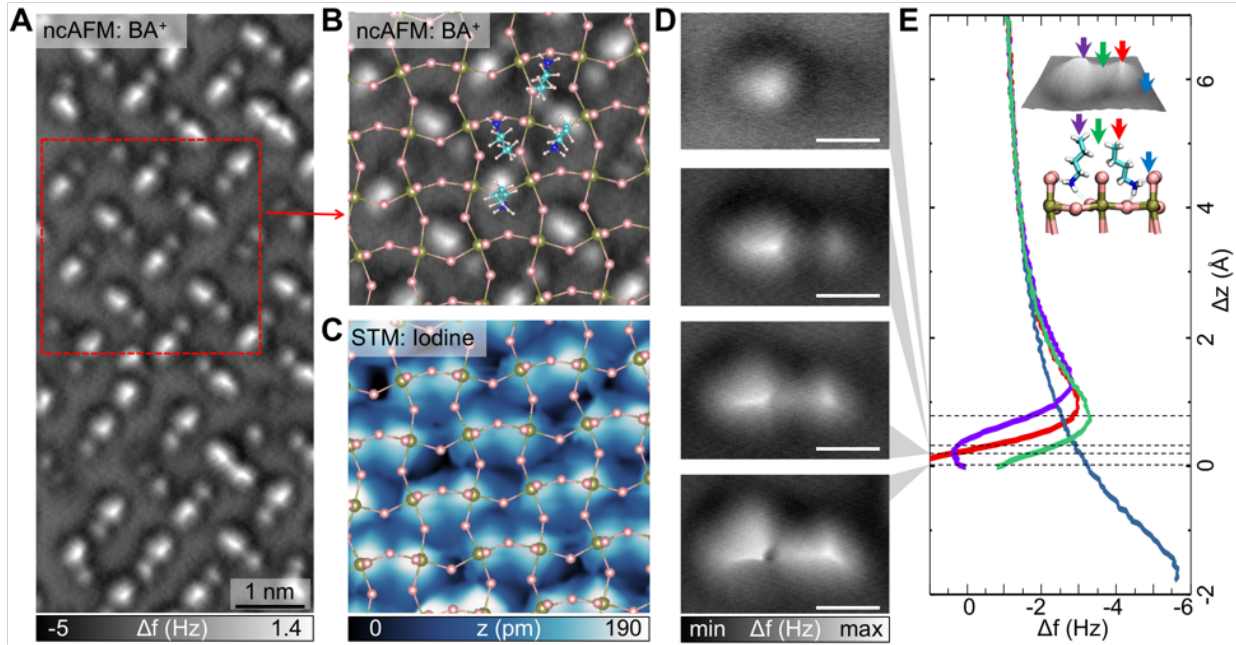

**Fig. 2. The STM and ncAFM imaging of organic and inorganic layers in few-layer RPP.** (**A**) Constant-height $\Delta f$ image. (**B**) zoom-in constant-height $\Delta f$ image of the surface region marked by a red rectangle in (A). (**C**) STM image of the same surface area as shown in panel B, superimposed with structure of the DFT-relaxed RPP lattice. The color-coding of elements: lead – green, iodine – pink, carbon – cyan, nitrogen – blue, hydrogen – white. (**D**) A set of constant-height ncAFM images collected at various tip-sample distances ($\Delta z$) over an individual pair of BA$^+$ cations. (**E**) $\Delta f$ versus $\Delta z$ curves acquired over the sites marked by color-coded arrows in the experimental 3D-rendered ncAFM image in inset (top) and side view of DFT-relaxed BA$^+$ pair structure in inset (bottom). The $\Delta z = 0$ is defined with respect to an STM setpoint of $V_s = 2$ V and $I = 15$ pA. The length of scale bar in ncAFM images shown in panel D is 0.3 nm.

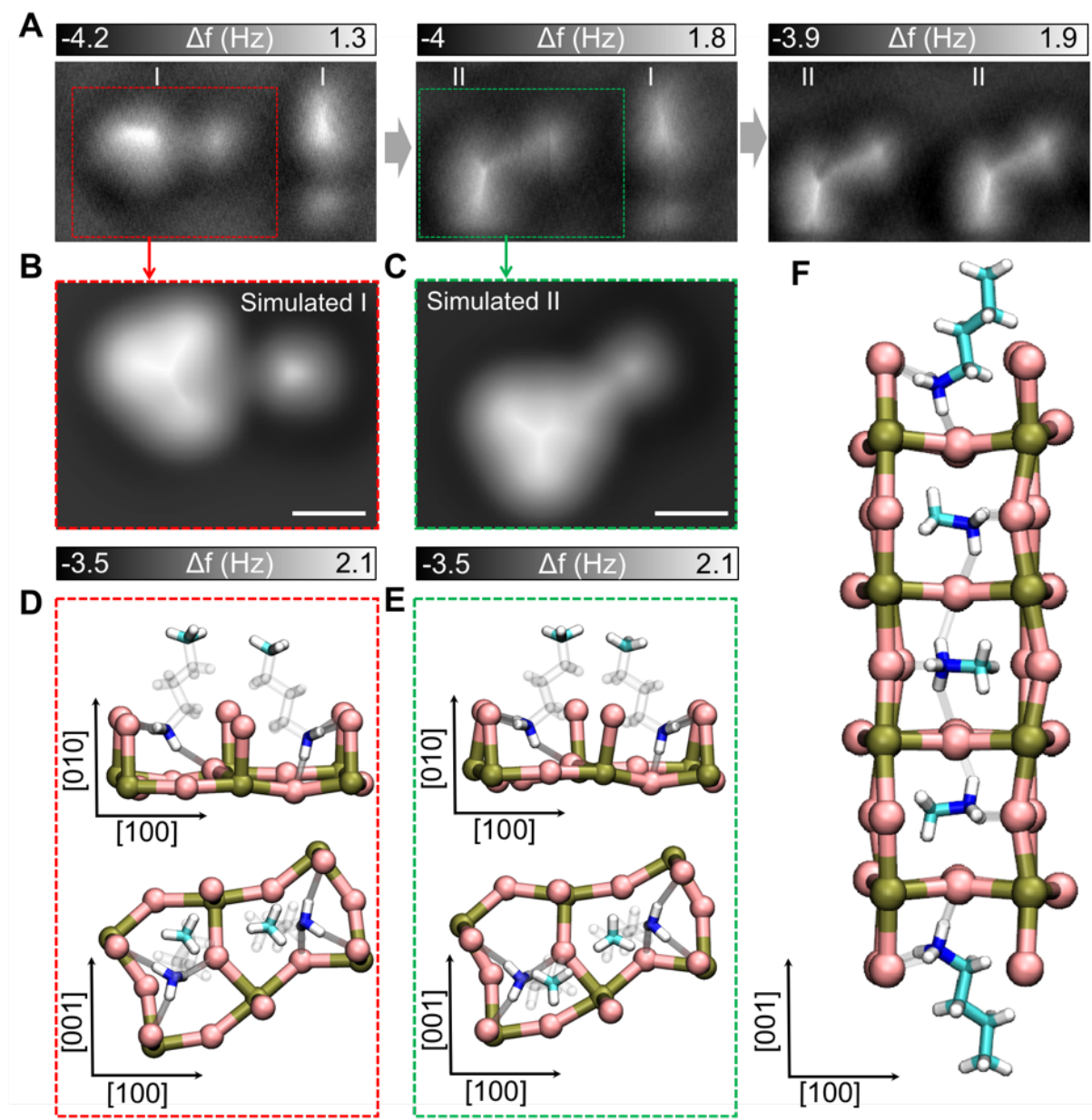

**Fig. 3. Probing the origin of the pairing of the BA$^+$ molecules.** (**A**) ncAFM images of the same surface area showing the tip-assisted transformation of the BA$^+$– pairs from type-I to type-II. (**B**) simulated ncAFM image of the "type-I" BA$^+$ pair. (**C**) simulated ncAFM image of the "type-II" BA$^+$ pair. (**D**) The side and top view of DFT-relaxed structure of "arrow-like" type-I BA$^+$-pair. (**E**) The side and top view of the DFT-relaxed structure of "T-like" type-II BA$^+$-pair. (**F**) side view of DFT-relaxed n=4 RPP slab structure. The color-coding of elements: Lead –

green, iodine – pink, carbon – cyan, nitrogen – blue, hydrogen – white. The length of scale bar in simulated ncAFM images shown in panels D and C is 0.3 nm.

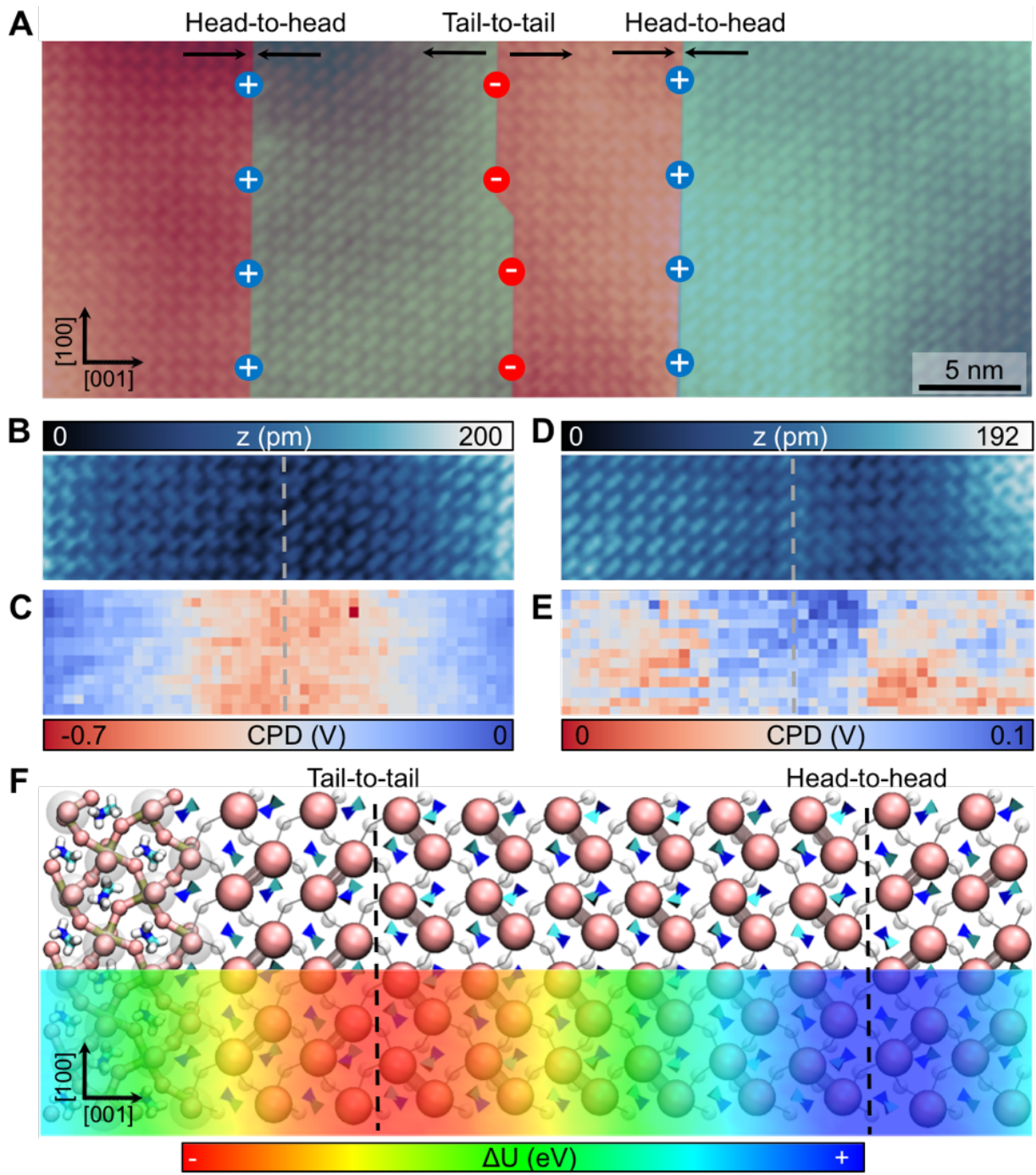

**Fig. 4| Imaging the domain structure and electrostatic potential across the twin-domain boundary. a,** Large-scale STM image of the RPPs reveals a twin–domain crystal composition. Distinct ferroelastic domains are color-coded by blue and red. **b**, **c** STM image (**b**) and corresponding 2D LCPD map (**c**) of the "head-to-head" twin-boundary. (**d**, **e**), STM image (**d**) and corresponding LCPD map (**e**) of the "tail-to-tail" twin-boundary. (**f**) The DFT-relaxed structure of the "head-to-head" and "tail-to-tail" twin–boundaries, superimposed with a surface electrostatic potential determined using lattice dipole model.

**Acknowledgements:** J. Lu acknowledges the support from MOE grants (MOE2019-T2-2-044 and R-143-000-B58-114). M. Telychko acknowledges support from A*STAR AME YIRG grant (Project No. A20E6c0098, R-143-000-B71-305). K. P. Loh acknowledges the support from MOE grant (MOE2019-T2-1-037). P. Jelinek acknowledges financial support from Praemium Academie of the Academy of Science of the Czech Republic, GACR project no. 20-13692X and CzechNanoLab Research Infrastructure supported by MEYS CR (LM2018110).

**Author contributions**: M.T and J.L. conceived and designed the experiments. M.T. performed all experiments related to the characterization of RPPs *via* combined STM/ncAFM measurements and data analysis. K.L. and I.A. performed synthesis of RPP crystals and their mechanical exfoliation onto Au surface, under supervision of L.K.P. S.A., J.M.-M. performed the DFT calculations under supervision of P.J. J.L, N.G, C.Z, contributed to the scientific discussion. The manuscript was written by M.T., L.J. and P.J with contributions of all co-authors.

**Competing interests**: Authors declare that they have no competing interests

**Data Availability:** All data are available in the main text or the supplementary materials.

**Supplementary Materials**

Materials and Methods

Figs S1 to S7

Supplementary Note 1: Additional STM and ncAFM images of the RPP.

Supplementary Note2: Electronic structure of RPP with insight into the bias-dependent STM contrast

Supplementary Note 3: KPFM measurements and site-specific variation of the local contact potential difference of the RPP.

Supplementary Note 4: Description of theoretical modelling of the electrostatic field above surface with domain boundary.

References (62-68)

# Supplementary Materials for

# Sub-angstrom Non-invasive Imaging of Atomic Arrangement in 2D Hybrid Perovskites

Mykola Telychko[1,†], Shayan Edalatmanesh[2,3†], Kai Leng[1,†], Ibrahim Abdelwahab[1,4], Na Guo[5], Chun Zhang[5], Jesús I. Mendieta-Moreno[2], Matyas Nachtigall[2], Jing Li[4], Kian Ping Loh[1,*], Pavel Jelínek[2,3*], Jiong Lu[1,4,*]

*Corresponding author. Email: chmluj@nus.edu.sg (J. Lu); jelinekp@fzu.cz (P. Jelínek); chmlohkp@nus.edu.sg (K. P. Loh)

**This PDF file includes:**

Materials and Methods

Figs S1 to S7

Supplementary Note 1: Additional STM and ncAFM images of the RPP.

Supplementary Note 2: Electronic structure of RPP with insight into the bias-dependent STM contrast

Supplementary Note 3: KPFM measurements and site-specific variation of the local contact potential difference of the RPP.

Supplementary Note 4: Description of theoretical modelling of the electrostatic field above surface with domain boundary.

**Reference** (*62-68*)

**Materials and Methods**

Sample preparation: The RPP single crystals were synthesized using three solid precursors—PbO, $C_4H_9NH_3I$ and $CH_3NH_3I$ *via* a temperature-programmed solution precipitation method, described in detail in our recent report (17). Afterwards, these crystals were transferred onto the flat Au surface via mechanical exfoliation in the inert glove-box environment.

The qPlus STM/ncAFM measurements: The STM and ncAFM experiments were performed in the UHV conditions at 4.4 K using a commercial Omicron LT STM/AFM machine. We used a qPlus sensor with a resonant frequency of $f_0$ = 28.5 kHz, quality factor of Q = 12000 and oscillation amplitude of $A$ = 100-120 pm, operated in frequency-modulation mode. All ncAFM images were collected in a constant-height mode. Stabilization parameters prior to feedback loop opening are indicated in the corresponding figures caption.

KPFM mapping measurements: Kelvin parabolas were acquired in a (50 × 12 pixel) grid covering "head-to-head" and "tail-to-tail" domain boundaries, with acquisition time of 30 seconds per pixel. The $V_{CPD}$ values were further extracted from each Kelvin parabola via parabolic fit using formula $\Delta f(V)=a(V_S-V_{CPD})2 + c$ (see Supplementary Note 3 for further details). These $V_{CPD}$ values were used to construct the 2D LCPD maps as shown in Fig. 4C, E.

Computational Methods: The DFT calculations were performed using the Fireball package[62]. All geometry optimizations and electronic structure analyses were performed using BLYP exchange-correlation functional (63,64) with D3 corrections (65) and norm-conserving pseudopotentials with a basis set of optimized numerical atomic-like orbitals (66). Systems were allowed to relax until the remaining atomic forces reached below $5 \times 10^{-2}$ eV Å$^{-1}$. Two slabs were used in this study: i) 2 × 2 unit cell containing 300 atoms; ii) specifically designed slab mimicking the domain boundaries with 3600 atoms in total. Brillouin reciprocal zone was sampled by a Monkhorst–Pack grid of 5 × 5 × 1 for the 2 × 2 slab and 1 k-point for the large domain boundaries slab. The theoretical AFM simulations were carried out using probe particle SPM code (46-48), with the Hartree potential obtained from the DFT calculations, probe particle stiffness of k = 0.7 N/m and Rc = 1.661 Å to simulate CO-functionalized tip.

**Supplementary Note 1: Additional STM and ncAFM images of the RPP.**

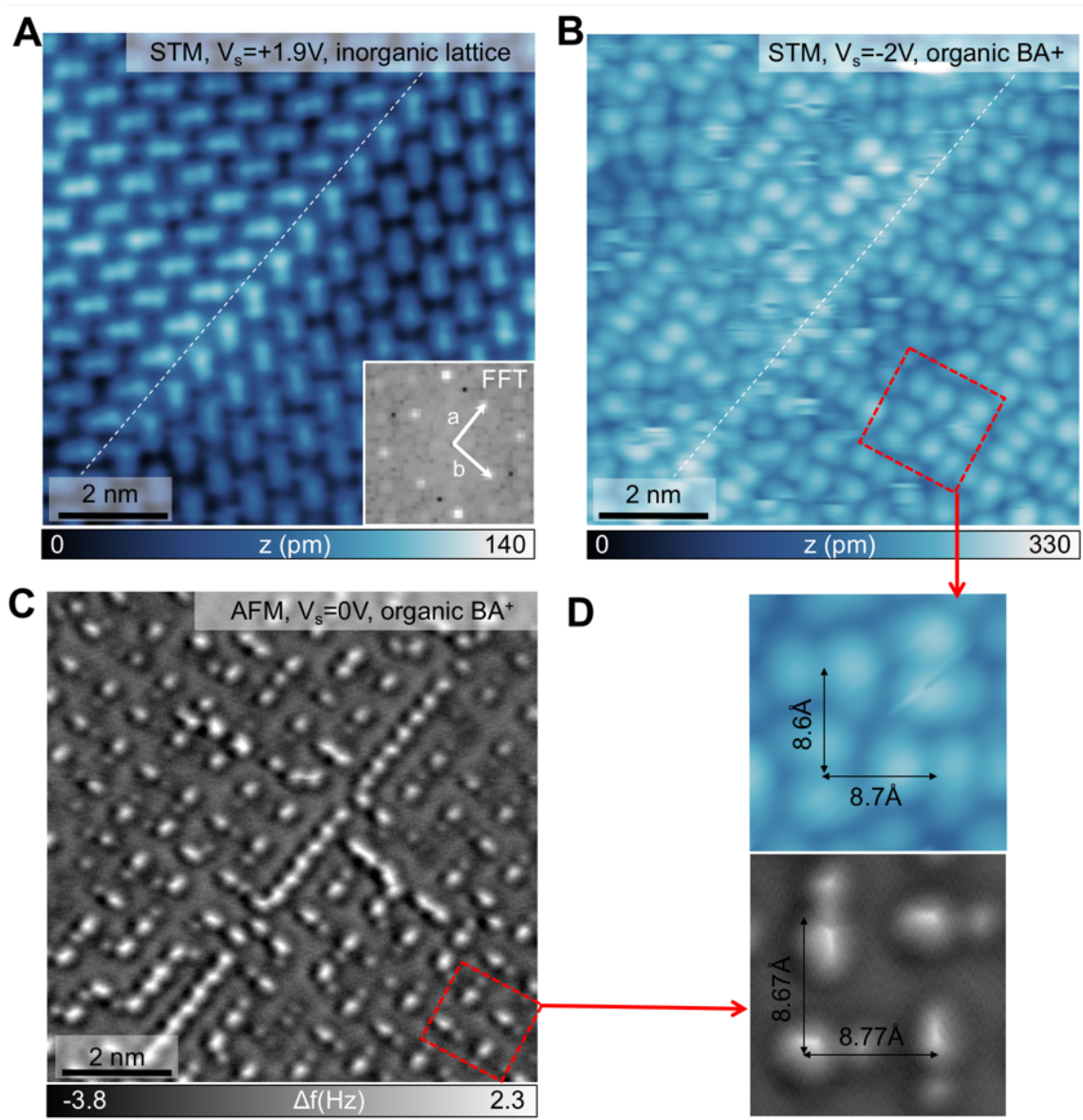

**Fig. S1. Combined STM and ncAFM imaging of the RPP.** (**A, B**) STM images of the RPP acquired over the same surface area at positive sample bias voltage ($V_s$ = +1.9 V) (panel A) and negative sample bias voltage ($V_s$ = -1.9 V) (panel B). White dashed line indicates the position of the twin-domain boundary. (**C**) ncAFM image of the same surface area. (**D**) zoom-in STM image (top panel) and ncAFM image (bottom panel) of the area indicated by dashed red rectangles in panels B and C respectively.

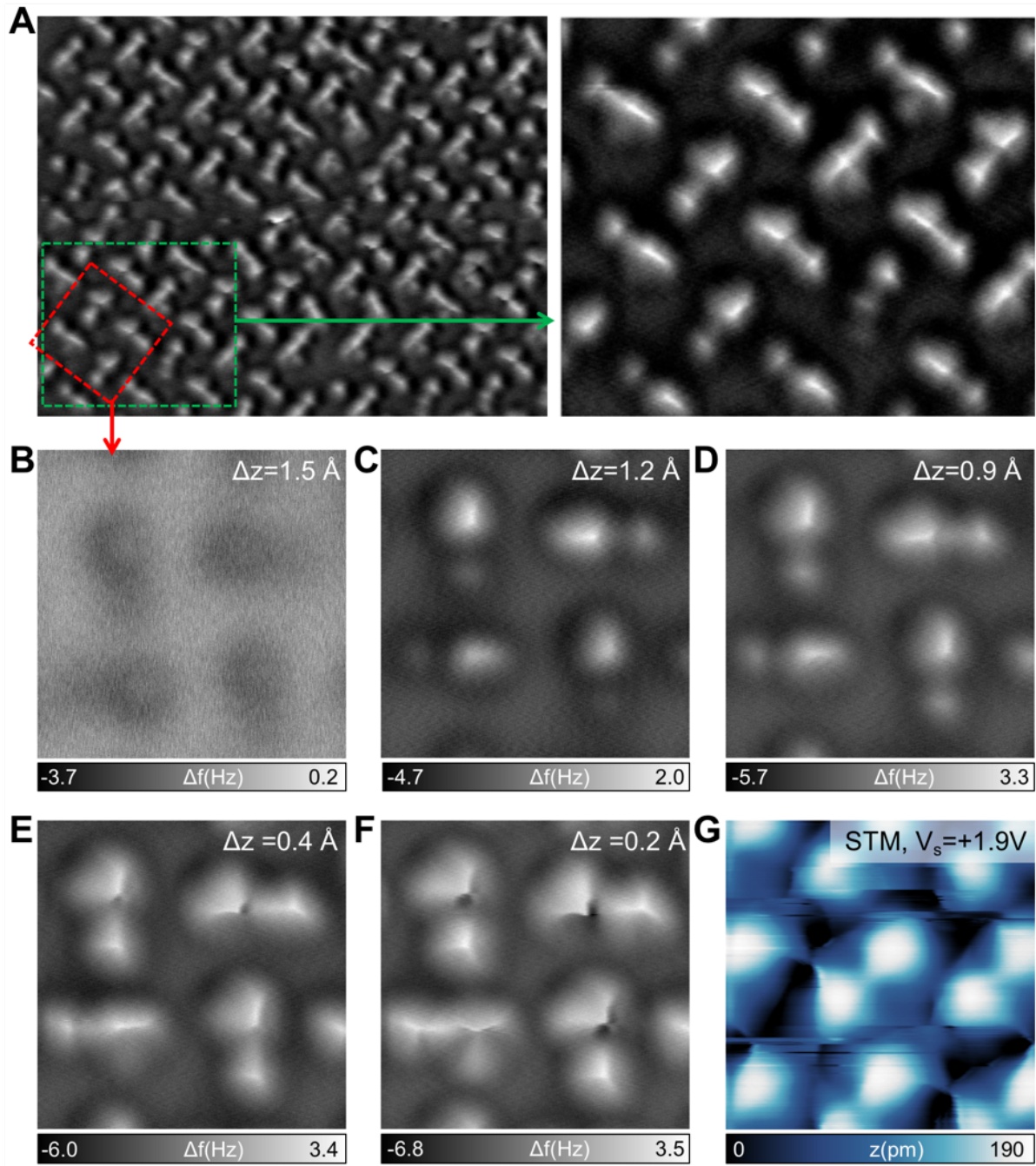

**Fig. S2. Large-scale ncAFM imaging of RPP.** (**A**) a representative large-scale constant-height ncAFM image of RPP flake. (**B-G**) a set of zoom-in ncAFM images (panels B-F) and STM image (panel G) collected within the surface area indicated by a red rectangle in panel A. ncAFM images were acquired at the tip-sample distance of $\mathit{\Delta z} = + 1.5$ Å (C), $\mathit{\Delta z} = + 1.2$ Å (D), $\mathit{\Delta z} = + 0.9$ Å (E), $\mathit{\Delta z} = +0.4$ Å (F), $\mathit{\Delta z} = + 0.2$ Å (G). The $\mathit{\Delta z} = 0$ is defined by the tip-sample distance corresponding to the tunneling set-point: $V_s = +2$ V and $I = 15$ pA.

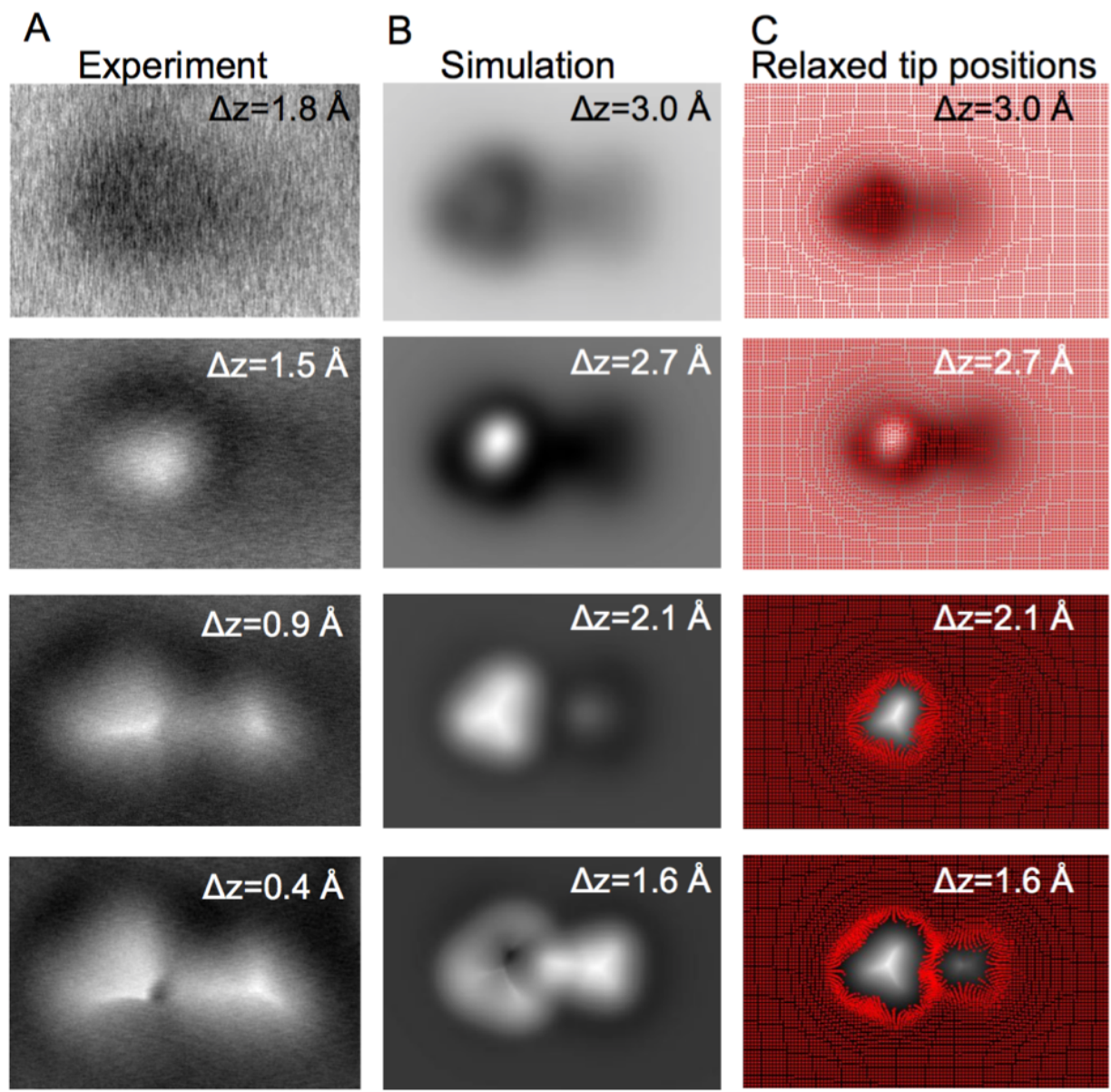

**Fig. S3. The evolution of the ncAFM contrast as a function of tip-sample distances.** (**A**) a set of constant-height ncAFM images collected at various tip-sample distances ($\Delta z$) over an individual pair of BA$^+$ cations. The $\Delta z$ value is indicated in the top-right corner of each panel. The $\Delta z=0$ is defined with respect to an STM set-point of $V_s = 2$ V and $I = 15$ pA. (**B**) A set of simulated ncAFM images using the probe-particle model. (**C**) same as panel B, but superimposed with relaxation trajectories of the CO molecule, where a mesh of red dots represents the *xy*-positions of the probe particle at given *z*-distances due to interaction with the sample.

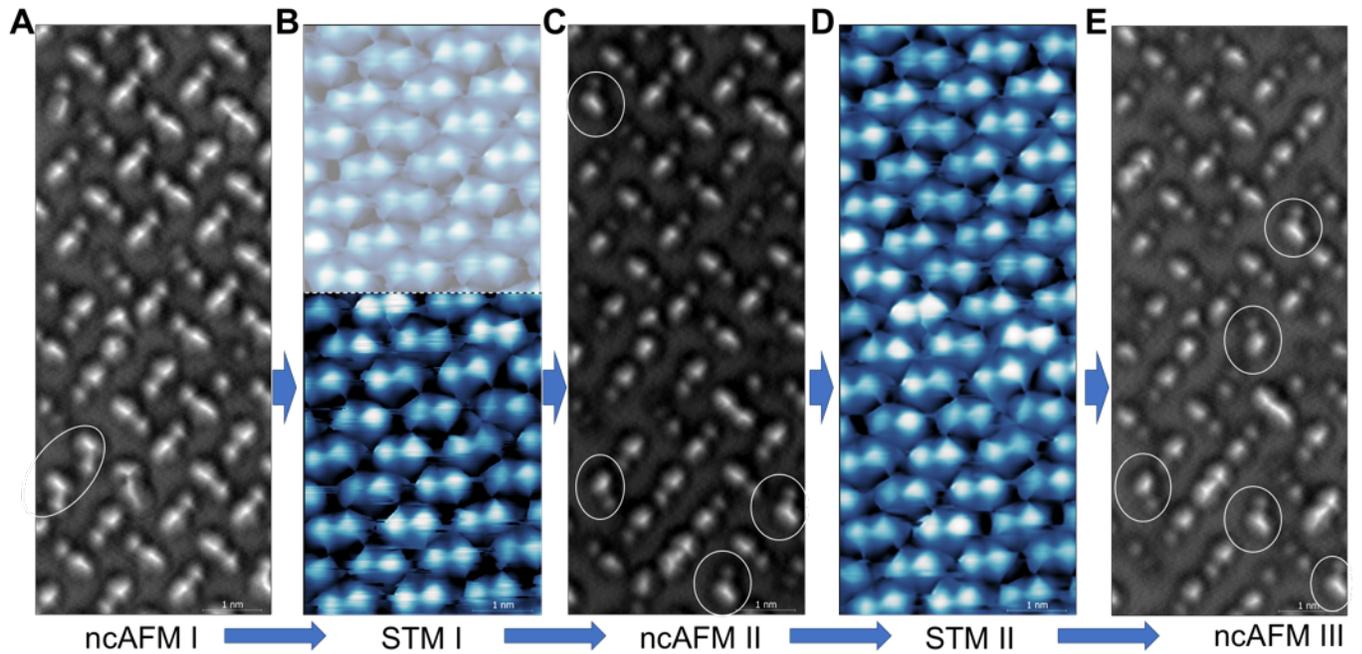

**Fig. S4. Subsequent ncAFM and AFM imaging.** Sequence of constant-height ncAFM (panels **A**, **C**, **E**) and STM images (panels **B**, **D**) of the same surface area. White circles denote type II $BA^+$ pairs. STM images were collected using sample bias of $V_s = +2$ V.

To illustrate invasive perturbation of the tunneling process onto $BA^+$ pairs, we acquired a sequence of ncAFM and STM images over the same surface area (Fig. S4). The ncAFM images show that $BA^+$ reconstruction undergoes remarkable changes, manifested by interconversion of $BA^+$ pairs from type I to type II (marked by white circles). In contrast, STM images collected at positive sample bias voltage reveal fully identical dimer-like pattern associated with the outermost I atoms, suggesting that tunneling process does not affect the underlying inorganic lattice.

**Supplementary Note2: Electronic properties of RPP with insight into the bias-dependent STM contrast**

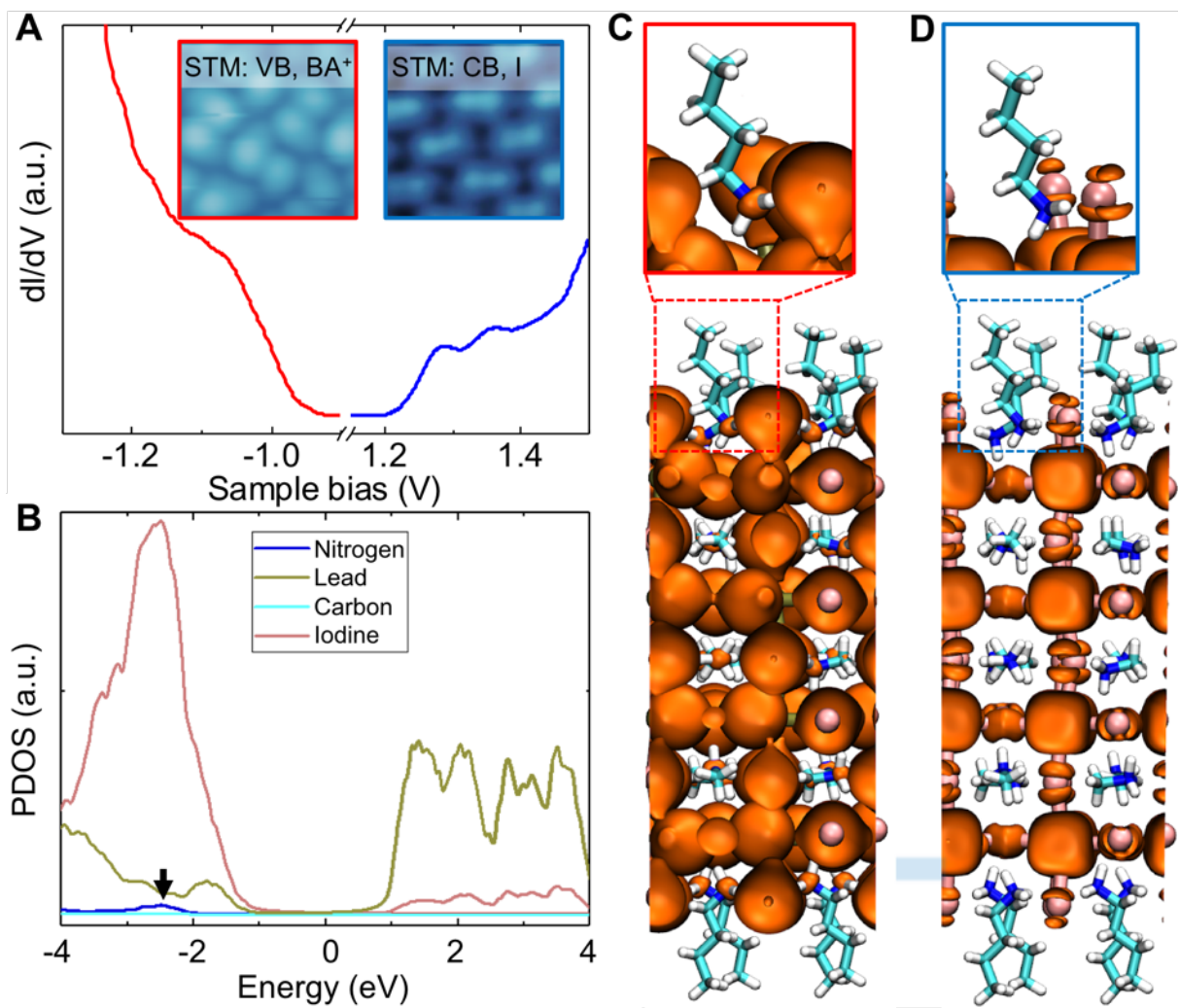

**Fig. S5. Electronic structure of RPP.** (**A**) A representative d$I$/d$V$ spectrum collected over the RPP flake. (**B**) calculated partial density of states of the $n$=4 RPP, normalized to the number of atoms in the unit cell. Black arrow indicates a non-negligible contribution of the BA$^+$ induced states in the valence band. (**C**, **D**) real-space distribution of the valence band (panel C) and conduction band states (panel D).

We probed electronic properties of the RPP by differential conductance spectroscopy (d$I$/d$V$) measurements (Fig. S5A). The representative d$I$/d$V$ spectrum collected over a RPP flake, exhibits a valence band maximum (VBM) at -1.06 eV and conduction band minima (CBM) at 1.11 eV.

The size of the electronic band gap is determined to be ~2.18 eV, which reasonably agrees with the calculated band gap (2.2 eV) (Fig. S5B) and previously reported ones (*67*). The calculated partial density of electronic states (PDOS) reveals that the conduction band (CB) of the RPP crystal is mainly composed of *6s* Pb and *5p* I electronic states (Fig. S5B). Therefore, tunneling current is dominated by the topmost I atoms, resulting in the dimer-like STM pattern (Fig. S1A, and inset of Fig. S5A) at positive sample bias voltage. In contrast, valence band (VB) of RPP, predominantly formed by *5p* I states, also contains a non-negligible contribution of the BA$^+$ induced states (indicated by red arrow in Fig. S5), which allows for the visualization of the on-surface BA$^+$ cations at negative sample bias voltages (Fig. S1B, inset of Fig. S5A).

**Supplementary Note 3: KPFM measurements and variation of the local contact potential difference over twin-boundaries.**

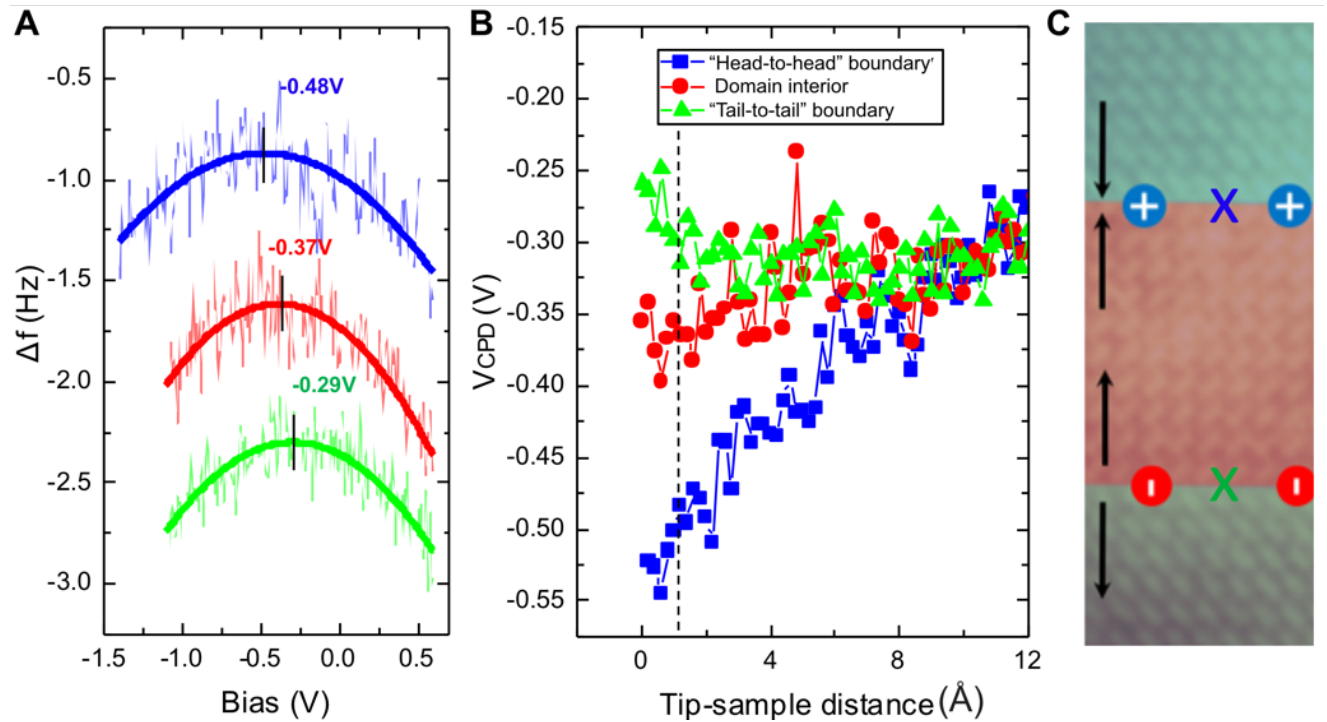

**Fig. S6. Kelvin probe force microscopy measurements.** (A) representative site-specific *Δf(V)* curves acquired within domain interior (red curve), over "head-to-head" twin-boundary (blue curve) and "tail-to-tail" twin-boundary (green curve). These *Δf(V)* curves were collected at tip sample distance of *Δz*=+1 Å with respect to *Δz*=0, defined by the tip-sample distance corresponding to the tunneling set-point: $V_s$ =+ 2.2 V and *I* = 15 pA. (**B**) evolution of the $V_{CPD}$

values extracted from the *Δf(V)* curves measured at different-tip sample distances. (**C**) a representative STM image showing the coexistence of "head-to-head" and "tail-to-tail" twin-boundaries. Crosses mark sites where *Δf(V)* curves were acquired.

To get insight into variations of the electrostatic potential across twin-boundaries, we conducted Kelvin probe microscopy measurements. Fig. S6A shows representative Kelvin parabolas (*Δf(V)* curves), collected at tip-sample distance of *Δz* = +1 Å over specific sites (see crosses in Fig. S6C), including (i) within domain interior (red curve), (ii) over "head-to-head" twin-boundary (blue curve) and (iii) over "tail-to-tail" twin-boundary (green curve). The local contact potential difference ($V_{CPD}$) values were further extracted from these *Δf(V)* curves *via* parabolic fitting using formula *Δf*=*a*(*V*-$V_{CPD}$)$^2$+*c*. We also plot the $V_{CPD}$ values versus different tip-sample distances (Fig. S7B).

At reduced tip-sample distance (*Δz* < +4 Å), we observe a clear variation of the $V_{CPD}$ over the twin-boundaries compared to the interior of the domain. Specifically, the $V_{CPD}$ over the "head-to-head" ("tail-to-tail") twin boundary is shifted to a lower (higher) value reflecting an increase (decrease) of the local work function (*56,68*), induced by a spatial variation of the surface electrostatic potential. The "head-to-head" and "tail-to-tail" boundaries are expected to host the corresponding positive and negative electrostatic potential as compared to the domain's interior.

**Supplementary Note 4: Description of theoretical modelling of the electrostatic field above surface with domain boundary**

The experimental KPFM contrast is driven by a variation of the electrostatic field in the *xy*-plane above the surface. To understand the role of individual molecular layers on the resulting electrostatic field at the domain boundaries in *xy*-plane above the surface, we established a simple theoretical model. In this model, we consider only the organic cations in periodic boundary conditions and leave aside the inorganic cage's potential since the latter only provides a background signal to the electrostatic field. Our model takes into account each of individual MA molecules as an electric dipole centered in the middle of the C-N bond, with the methyl (nitrate) group as the end with negative (positive) charge. For the BA molecules, we proceeded similarly and used only their electric dipole components in the *xy*-plane. The corresponding dipole moments of both MA⁺ and BA⁺ were calculated for freestanding molecules using the DFT method (MA⁺=2.13 D, BA⁺=3.14 D in $xy$ = 1.74 D). Orientation of individual dipole moments was adopted from the fully optimized atomic slab calculation of the domain boundary obtained from the DFT supercell calculation. The resulting variation of the electrostatic potential in $xy$-plane 10 Å above the topmost surface (Fig. 4F) was calculated on numerical grid summing the contribution of each electric dipole moment's potential using a simple formula as below:

$$\phi(R) = \frac{1}{4\pi\varepsilon_0} \frac{P \cdot \hat{R}}{R^2}$$

where $P$ is the dipole moment, $R$ is the position vector relative to the center of the dipole, $R = r - \frac{r_+ + r_-}{2}$, with $r_-$ ($r_+$) being the position of the nitrate (methyl) group. The density of the numerical grid in the xy-plane was 35x105 (sampling points). In this way, the contribution of individual molecular layers to the variation of the electrostatic potential above the surface can be easily disentangled as shown in Fig. S7. Note that a variation observed in a map of total electrostatic potential shown in Fig S7A matches a variation of the electrostatic potential obtained from DFT very well.

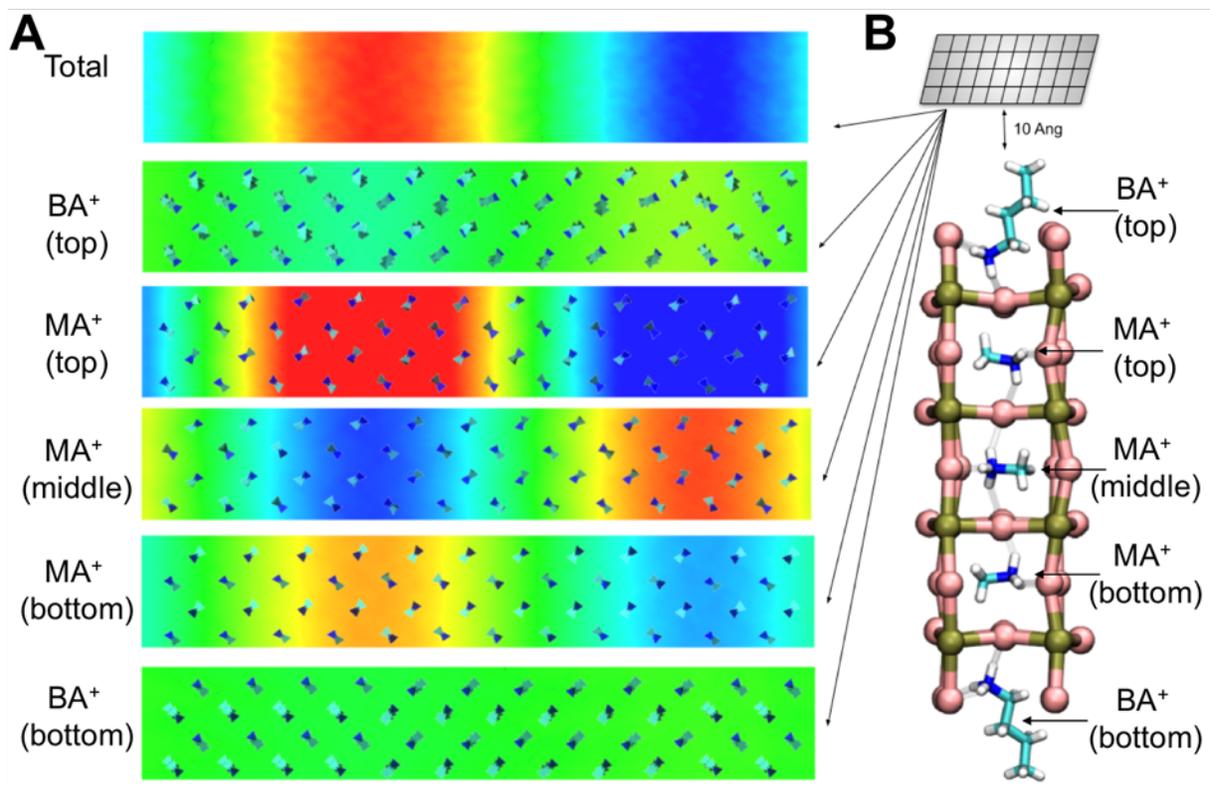

**Fig. S7. Theoretical modelling of the electrostatic field.** (**A**) variation of electrostatic potential in *xy*-plane 10 Å above the surface determined using an effective lattice dipole model across BA$^+$ and MA$^+$ planes. (**B**) schematic view of the procedure how the variation of the electrostatic potential in *xy*-plane 10 Å above the surface was calculated on a numerical grid summing the contribution of each electric dipole moment's potential in periodic boundary conditions